\documentclass[a4paper,11pt]{article}
\usepackage{pos}
\usepackage{multirow}
\usepackage{subfigure}

\title{The pole counting rule and XYZ states}
\author*[a]{Hao Chen}
\author[b]{Hong-Rong Qi}
\affiliation[a]{Department of Physics and Key Laboratory of Nuclear Physics and Technology, Peking University,\\
  209 Chengfu Road, Beijing, China}

\affiliation[b]{Department of Engineering Physics, Tsinghua University,\\
30 Shuangqing Road, Beijing, China}

\emailAdd{haochen0393@pku.edu.cn}
\emailAdd{qihongrong@tsinghua.edu.cn}
\abstract{The pole counting rule is a powerful and model-independent method to distinguish a confining state from a hadronic molecule. It has been applied to the explorations of $X(6900)$, $X_1(2900)$ as well as $Z_c(3900)$, $X(3872)$, $X(4660)$, etc.. For $X(6900)$, both a confining state and a molecular state are not excluded, because lacking of enough data. For $X_1(2900)$, the analysis shows that it should be a $\Bar{D}_1 K$ molecule, with $J^P=1^-$ and an iso-singlet interpretation is much more favorable. Finally, it is noted that almost all XYZ particles with exotic quantum numbers can be interpreted as hadronic molecules. The $X(3872)$ is, however, more like a charmonium. }

\FullConference{%
   10th International Workshop on Chiral Dynamics \\
    15-19 Nov. 2021 \\
   Institute of High Energy Physics, CAS, Beijing, China
}

\begin{document}
\maketitle

\section{Introduction}
In the past 20 years, many exotic hadrons (also known as ``XYZ" states) have been observed. These states cannot be explained by ordinary quark model for mesons and baryons which have valence quarks of $\bar{q}q$ and $qqq$, respectively. So, a natural idea could be that these states have multiquark constituents, i.e., they may have valence quarks of $\bar{q}\bar{q}qq$, $\bar{q}qqqq$. An explanation for these multiquark states is that they are compact or elementary states, whose building blocks are quarks and anti-quarks and they interact with each other by exchanging gluons. However, there is another explanation that these XYZ states are molecules of ordinary hadrons. From this point of view, the building block of these states are ordinary mesons or baryons and they interact with each other by exchanging color neutral forces. On the one hand, multiquark states are not forbidden by our present understanding of QCD, the underlying theory of strong interaction and this indicates that the compact state explanation is possible. On the other hand, these states are always observed near the thresholds of two-hadron channels which indicates that the hadronic molecule explanation is also possible. So, it is important to distinguish hadronic molecules from compact states and this has become one of the most active topics in hadron physics.

Many methods are proposed to do this task, for example, QCD sum rule, quasi-BS equation, diquark model, etc~\cite{Hosaka:2016pey,Brambilla:2019esw}. In this paper, the pole counting rule~(PCR), a model-independent method will be introduced and some applications on XYZ states such as $X(3872),~Z_c(3900),~X(4660)$ as well as $X(6900),~X_1(2900)$ will be reviewed. For $X(3872)$, the result from PCR suggests that it is more like a charmonium, i.e., not a hadronic molecule~\cite{Zhang:X3872,Meng:X3872}. The same situation also happens on $X(4660)$~\cite{Cao:X4660}. As for $Zc(3900)$ and $X_1(2900)$, they are regarded as hadronic molecules of $\bar{D}^* D$~\cite{Gong:Zc3900,Gong:Zc3900_2} and $\bar{D}_1 K$~\cite{Chen:X2900}, respectively. Furthermore, due to lack of experimental data on $X(6900)$, the fit results which are regarded as compact states as well as the results which are regarded as hadronic molecules according to PCR have the same fit goodness. So, one cannot confirm the nature of $X(6900)$ from current data~\cite{Cao:X6900}.

This talk will be organized as follows, firstly, the statements of PCR will be briefly reviewed. Then, the applications of PCR on $X(3872),~Z_c(3900),~X(4660)$ as well as $X(6900),~X_1(2900)$ will be introduced. Finally, some conclusions will be made. 

\section{The pole counting rule}
In this section, the statements of the pole counting rule will be introduced briefly. This method was proposed by D. Morgan in 1992~\cite{Morgan:1992ge}. Consider an $S$-wave scattering amplitude:
\begin{equation}
    T=(M-i k)^{-1},
\end{equation}
where $k$ is the momentum of scattering particles, and the pole position is determined by $M-ik=0$. If it is a potential scattering, then $M$ can be written in the form of effective range expansion:
\begin{equation}\label{M_effective}
    M=-\frac{1}{a_{0}}+\frac{r_{0}}{2} k^{2}+O\left(k^{4}\right), 
\end{equation}
where $a_0$ is the scattering length and $r_0$ is the effective range. Substitute (\ref{M_effective}) into the equation $M-ik=0$, a quadratic equation with respective to pole position is obtained. Two roots of this equation satisfy:
\begin{equation}\label{two_root_for_molecule}
    \left|k_{1}\right|+\left|k_{2}\right| \geq \frac{2}{\left|r_{0}\right|},
\end{equation}
where effective range of strong interaction usually takes the value of about (200~MeV)$^{-1}$, so the r.h.s. of the inequality~ (\ref{two_root_for_molecule}) is a quite large quantity. On the one hand, in order to satisfy this inequality, there is only one pole can appear around the threshold and another pole should located far enough from the threshold to make the l.h.s. of (\ref{two_root_for_molecule}) big enough. On the other hand, the potential scattering scenario means that the pole is dynamically generated, i.e., it represents that the structure should be regarded as a hadronic molecule.  

Moreover, if the pole hardly couples to the hadronic channel in which the structure is observed, then the pole will behave more like a CDD pole in the N/D method. In other words, the pole is more "elementary" and cannot be generated from hadron interactions. It should be regarded as a compact state. In this situation, $M$ can be written as,
\begin{equation}
    M=\frac{k^{2}-k_{0}^{2}}{g^{2}},
\end{equation}
two roots satisfy,
\begin{equation}\label{two_root_for_CDD}
    \left|k_{1}\right|+\left|k_{2}\right| \geq g^{2}.
\end{equation}
If the coupling constant $g$ is small, two poles can be found around the threshold in this case.

So, the pole counting rule provides a model-independent method to distinguish hadronic molecules from compact states. In the $S$-wave scattering, if there is only one pole near the threshold, then the scattering state should be regarded as a hadronic molecule. If there are two poles around the threshold, the scattering state should be regarded as a compact state. Usually, the poles will be searched for in the complex energy plane, but the criteria of molecular states and compact states are the same. It is noted that if there are more than one couple channels, the poles will have imaginary parts since the resonance can decay into lower channels and the pole positions will move to the complex energy plane. Furthermore, the PCR is still a qualitative method, and the efforts for a quantitative version can be found in~\cite{Xiao:2016mon}.

\section{Applications}
In this section, the applications of PCR on XYZ states such as $X(3872),~Z_c(3900),~X(4660)$ as well as $X(6900),~X_1(2900)$ will be introduced. 
\subsection{$X(3872)$}
$X(3872)$ was firstly observed in 2003 in the $J/\psi \pi^+ \pi^-$ final state~\cite{Belle:X3872} and then it was observed in $\bar{D}^0 D^0 \pi^0$ final state and near the threshold of $\bar{D}^{0*} D^0$~\cite{BaBar:X3872}. So, it is natural to think $X(3872)$ to be a molecule of $\bar{D}^* D$ or a compact state, e.g., a charmonium or a tetraquark state. 

In order to study the nature of $X(3872)$, Ref.~\cite{Zhang:X3872} considered the coupling between $X(3872)$ and $\bar{D}^{*0} D^0, ~D^{*-}D^+$, as well as $J/\psi \pi^+ \pi^-,~J/\psi \pi^+ \pi^- \pi^0$. Using Flatt\'e-like parametrization, the data sets corresponding to the final states of $\bar{D}^{*0} D^0$, $\bar{D}^0 D^0 \pi^0$, $J/\psi \pi^+ \pi^-$ were fitted. During the fitting, the parameter $\Gamma_c$ which represents the decay width of some modes far away from the threshold such as $\psi^{\prime}\gamma$ can provide important information on the nature of $X(3872)$. If $X(3872)$ is a charmonium, then its most important hidden-charm decay mode is the inclusive light hadronic decay. In this case, $\Gamma_c\sim \mathcal{O}$(1)~MeV~\cite{Meng:2007cx}. But if $X(3872)$ is a $\bar{D}^{*0} D^0$ molecule, the most important hidden-charm decay mode is decaying to $\chi_{cJ}\pi\pi$. In this case, $\Gamma_c$ is expected to be smaller than $\Gamma_{J/\psi \pi\pi}\sim$0.05~MeV~\cite{Fleming:2008yn}. 

Besides $\Gamma_c$, the parameter $\mathcal{B}$ representing the branch ratio of $B\to X(3872)K$ is also important. From the fit results, it is found that if $\mathcal{B}$ takes a reasonable value, there always exist two poles near the threshold and $\Gamma_c$ will be at $\mathcal{O}(1)$~MeV which supports the charmonium statement. However, the fit result corresponding to pure molecular state where there is only one pole near the threshold can also be found. But in this case, the value of $\mathcal{B}$ is too large to be realistic and $\Gamma_c$ is of  $\mathcal{O}(1)$~MeV which is inconsistent with the analyses in Ref.~\cite{Fleming:2008yn}. So, with the help of PCR, it is suggested in Ref.~\cite{Zhang:X3872} that $X(3872)$ is more like a charmonium rather than a pure hadronic molecule, but the coupling to $\bar{D}^{*0}D^0$ is also important, since the pole of $X(3872)$ is located around the threshold of $\bar{D}^{*0}D^0$.

\begin{figure}[htpb]
    \centering
    \includegraphics[scale=0.4]{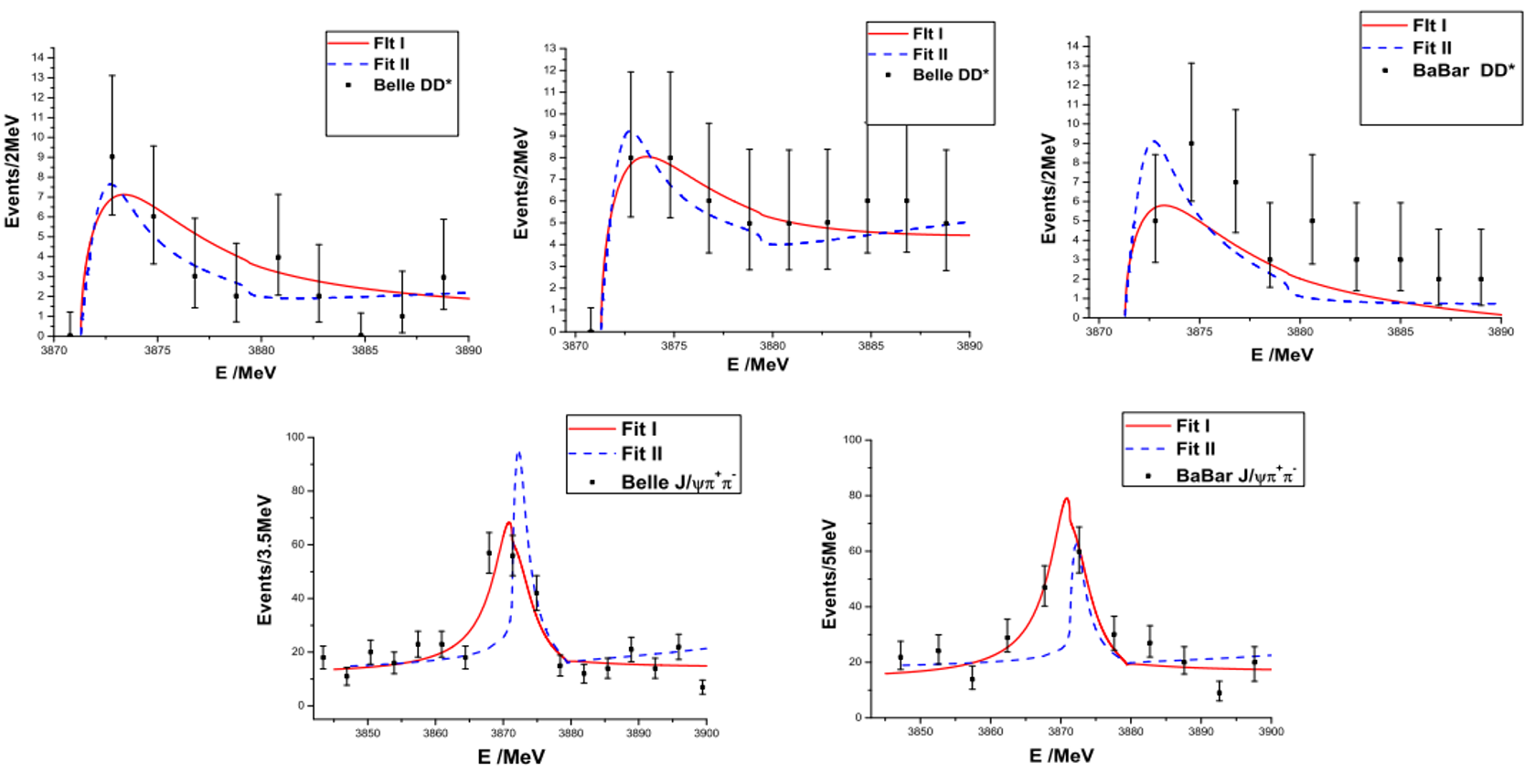}
    \caption{Fitting results of two methods in Ref.~\cite{Meng:X3872}.}
    \label{fig:X3872}
\end{figure}

Furthermore, in Ref.~\cite{Meng:X3872} the dynamically generated $X(3872)$ which is produced by $\bar{D}^* D$ bubble chain and the explicitly introduced $X(3872)$ as well as the competition between these two schemes, were under consideration. It is found that a pure explicitly introduced $X(3872)$~(called Fit I in Ref.~\cite{Meng:X3872}) can fit the data better than a pure dynamically generate $X(3872)$~(called Fit II in Ref.~\cite{Meng:X3872}) as shown in Fig.~\ref{fig:X3872}. As the results, Fit I gives more than one pole around the threshold of $\bar{D}^* D$, which indicates that $X(3872)$ is more like a charmonium in accordance with the PCR.
%. According to the PCR, it suggests that $X(3872)$ is more like a charmonium rather a hadronic molecule. 

\subsection{$Z_c (3900)$}
$Z_c(3900)$ is another well known exotic state besides $X(3872)$. It was firstly observed in the $J/\psi \pi$ final state by BESIII~\cite{BESIII:3900}, with the mass $(3899.0\pm 3.6)$~MeV and the width $(46 \pm 10)$~MeV determined by Breit-Wigner fit. Soon after, $Z_c (3900)$ was also observed in the $\bar{D}^* D$ final state~\cite{BESIII:3900II} where the peak is very close to the threshold. So, a natural assumption is regarding it as the hadronic molecule of $\bar{D}^* D$, but the scenario that $Z_c(3900)$ is a compact state, e.g. a tetraquark state cannot be excluded.

Ref.~\cite{Gong:Zc3900} constructed the amplitudes of processes $e^+ e^- \to J/\psi \pi \pi,~ \bar{D}D^* \pi,~h_c \pi\pi$ and fitted the invariant mass spectrum of $\pi\pi,~J/\psi \pi,~\bar{D}^* D$ simultaneously. Two mechanisms are involved, one is dynamically generated by $\bar{D}^* D$ contact interaction~(corresponding to Fit I) and another is exchanging explicit $Z_c$ in $s$ channel~(corresponding to Fit II). These two mechanisms are quite different, but they can lead to the same result as shown in Ref.~\cite{Gong:Zc3900}. Only one pole can be found around the threshold of $\bar{D}^* D$ in both Fit I and Fit II. In Fit I, the pole is located at $(3.880- 0.004i)~$GeV and in Fit II, the pole is located at $(3.979- 0.001 i)~$GeV. According to the PCR, it can be concluded that $Z_c (3900)$ is a hadronic molecule of $\bar{D}^* D$. 

\subsection{$X(4660)$}
The PCR was also applied to study the properties of $X(4660)$. $X(4660)$ was firstly observed by Belle~\cite{Belle:4660} in 2007. Later, it was observed in the final state $\Lambda_c \bar{\Lambda}_c$~\cite{Belle:4660II} and the peak is near the threshold of $\Lambda_c \bar{\Lambda}_c$. However, in 2017, BESIII released new data on $X(4660)$ and it was shown that the new data conflicted with previous data. 

As shown in Ref.~\cite{Cao:X4660}, this conflict between the old and the new data can be explained by a virtual pole near the $\Lambda_c \bar{\Lambda}_c$ threshold rather than the threshold enhancement of electromagnetic force. The $\Lambda_c \bar{\Lambda}_c$ contact interactions, as well as a $X(4660)$ Breit-Wigner resonance are combined to explain the odd line-shape. The cross sections of $e^+ e^- \to \Lambda_c \bar{\Lambda}_c$ and $e^+ e^-\to \psi^{\prime}\pi\pi$ are fitted. A K-matrix formula which mixes the contact interaction and $s$ channel $X(4660)$ exchange is used to describe $\Lambda_c \bar{\Lambda}_c$ final state interactions :
\begin{equation}
    A_K = \frac{1}{1-i\rho K}=\frac{1}{1-i\rho(T_X+T_{\text{ci}})},
\end{equation}
where $\rho$ is the two-body phase space factor of $\Lambda_c \bar{\Lambda}_c$ and $T_X$, $T_{\text{ci}}$ are the tree diagrams of $s$ channel $\Lambda_c \bar{\Lambda}_c\to X(4660) \to \Lambda_c \bar{\Lambda}_c$ and $\Lambda_c \bar{\Lambda}_c$ contact vertex as final states interactions. From the pole analyses, there is a stable virtual pole with mass of $M_{V}=(4.566\pm 0.003)$~GeV. Furthermore, $X(4660)$ is a compact state, if exists, since there are two nearby poles.

\subsection{$X(6900)$ and $X(4260)$}
In 2020, LHCb observed a structure around 6.9~GeV, named as $X(6900)$, in the di-$J/\psi$ invariant mass spectrum~\cite{LHCb:6900}. This state is the first observed state which has four (anti-)~charm quarks as valence quarks. Since it is difficult to produce $\bar{c}c$ from the vaccum, so $X(6900)$ is hardly a excited charmonium. For this reason, most studies suggested that $X(6900)$ is a tetraquark state~\cite{Chenhx:6900,Deng:6900} or a hadronic molecule~\cite{Gongc:6900,Liang:6900}. In Ref.~\cite{Cao:X6900}, Flatt\'e-like parametrization was involved to describe the structure around 6.9~GeV as well as a higher structure around 7.2~GeV, called $X(7200)$. Using this parametrization, the di-$J\psi$ invariant mass spectrum was fitted and the nature of these structures were analysed according to the PCR.

%The Flatt\'e-like parametrization involved in Ref.~\cite{Cao:X6900} is: 
%\begin{equation}\label{components}
%    \begin{aligned}
%        &\mathcal{M}_{i}=\frac{g_{i}n_{i1}(s) e^{i \phi_{i}}}{s-M_{i}^{2}+i M_{i} \sum_{j} \Gamma_{i j}(s)},
       %&\mathcal{M}_{\rm NoR}=c_0 e^{c_1(\sqrt{s}-2 m)} \sqrt{\frac{s-4 m^{2}}{s}}, 
%\end{aligned}
%\end{equation}
%where $g_i$ is the coupling constant and $\phi_i$ is the interference phase. $M_i$ is the line-shape masses of resonances and $i=1,~2,~3$ is for the structures at $6.5~$GeV, 6.9~GeV, 7.2~GeV, with respectively. $\Gamma_{ij}$ corresponds to  the partial width of the $j$-th couple channel on the $i$-th pole and $n_{ij}$ is the form factor of this couple channel:
%\begin{equation}
%\label{eq:Gamma}
%    n_{ij}(s)=\left(\frac{p_{ij}}{p_0}\right)^l F_l(p_{ij}/p_0), \quad \Gamma_{ij}(s)=g_{ij}\rho_{ij}(s)n_{ij}^2(s),
%\end{equation}
%where, $l$ is the orbital angular momentum in channel $j$, $p_{ij}$ is the center-of-mass momentum of one daughter particle of channel $j$ for two body decays.  $p_0$ denotes a momentum scale,
%$g_{ij}$ is a coupling constant,  and $\rho_{ij}(s)=2p_{ij}/\sqrt{s}$, is the phase space factor. $F_l$ is the so called Blatt-Weisskopf factor~\cite{Blatt-Fl}.

\begin{figure}[htpb]
    \centering
    \subfigure[]{
    \includegraphics[scale=0.20]{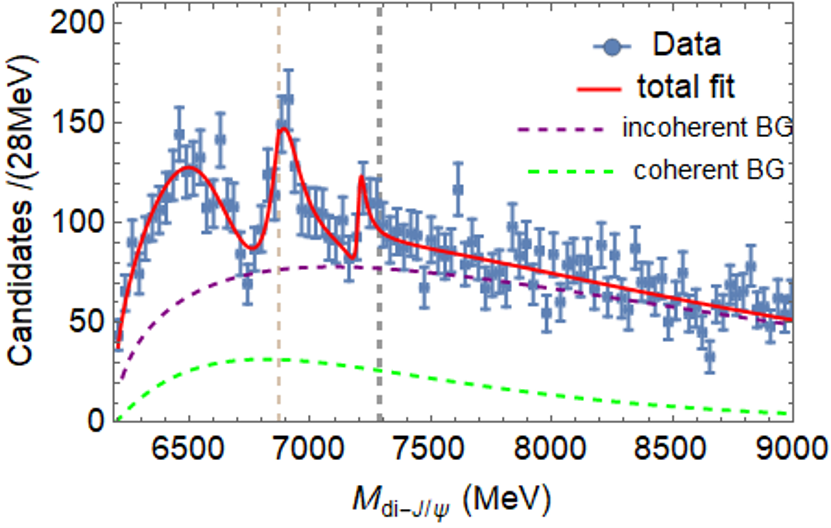}}
    \subfigure[]{
    \includegraphics[scale=0.20]{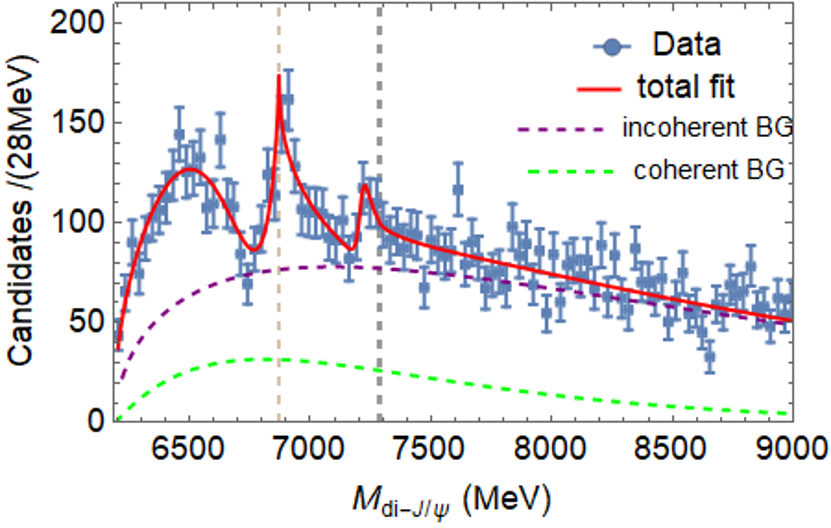}}
    \caption{Fit results of case I in S-S coupling in Ref.~\cite{Cao:X6900}: (a): compact state, (b): molecular state. }
    \label{fig:X6900}
\end{figure}

Three types of fits are carried out in Ref.~\cite{Cao:X6900} called S-S coupling, P-P coupling, S-P~(or P-S) coupling, where the first and the second character represent the angular momentum quantum number of channels coupled to $X(6900)$ and $X(7200)$, respectively. From the results, it is found that both the solutions of compact states and the solutions of molecular states can be found and these two types of solutions have almost the same fit goodnesses~(an example is shown in Fig.~\ref{fig:X6900}). As an example, the pole positions of S-S couplings are listed in Tab.~\ref{tab:sscoupling}. Here, the solution I stands for compact state and solution II stands for hadronic molecule. Three cases for the $X(6900)$ coupling are under consideration:
Case I: $J/\psi J/\psi$ and $J/\psi \psi(3770)$,     
Case II: $J/\psi J/\psi$ and $J/\psi \psi_2(3823)$,
Case III:$J/\psi J/\psi$ and $J/\psi \psi_3(3842)$. 
For $X(7200)$, the $J/\psi J/\psi$ and the near-threshold $J/\psi$ $\psi(4160)$ channels are used in the couple channel analysis. 

\begin{table}[htpb]
    \centering
\begin{tabular}{ccccc}
\hline
\hline 
& Case & State & Sheet II & Sheet III \\
\hline 
\multirow{6}{*}{Sol. I} & \multirow{2}{*}{I}& $X(6900)$ & $6885.4-68.0 i$ & $6874.4-80.0 i$ \\
& & $X(7200)$ & $7202.2-16.6 i$ & $7187.1-18.0 i$ \\
&\multirow{2}{*}{II}& $X(6900)$ & $6947.6-172.0 i$ & $6810.4-274.0 i$ \\
& & $X(7200)$ & $7220.8-31.0 i$ & $7220.8-31.0 i$ \\
&\multirow{2}{*}{III}&$X(6900)$& $6845.2-117.0i$ & $6789.2-138.0i$ \\
& & $X(7200)$ & $7221.9-28.0 i$ & $7221.9-28.0 i$ \\
\hline 
\multirow{6}{*}{ Sol. II } &\multirow{2}{*}{I}& $X(6900)$ & $6937.9-97.0 i$ & $6527.3-323.0 i$ \\
& & $X(7200)$ & $7210.7-27.5 i$ & $7037.3-47.5 i$ \\
&\multirow{2}{*}{II}& $X(6900)$ & $6933.9-111.0 i$ & $6443.8-275.0 i$ \\
& & $X(7200)$ & $7218.9-24.0 i$ & $7067.9-41.5 i$ \\
&\multirow{2}{*}{III}& $X(6900)$ & $6933.3-113.0 i$ & $6452.3-275.0 i$ \\
& & $X(7200)$ & $7221.9-23.0 i$ & $7073.7-41.0 i$ \\
\hline
\hline
\end{tabular}
    \caption{Summary of pole positions~(written in MeV) for  $S-S$ couplings. In practice, it is adopted that if a pole is more than three times of line-shape width away from the threshold, it can be seen as a faraway pole. So, for every case, there is only one nearby pole in solution II. The Riemann sheets are defined by changing signs of the phase space factors for the first and the second couple channels where $(+,+),~(-,+),~(-,-),~(+,-)$ stand for the sheet I, II, III, IV, respectively. }
    \label{tab:sscoupling}
\end{table}

According to the pole analysis in Ref.~\cite{Cao:X6900}, the nature of $X(6900)$ as well as $X(7200)$ cannot be determined using current data. So, more data are required to study their properties better.

It is worthwhile to mention that in Refs.~\cite{Dai:4260,Cao:4260}, similar methods are applied to study $X(4260)$, another exotic state. The couplings to two nearby channels, $\omega \chi_{c0},~D_{s}^* \bar{D}_s^*$ are carefully considered. The results shows that with a reasonable production widths of $e^+ e^- \to X(4260)$, the couplings of $X(4260)$ to these channels are not strong enough to form a hadronic molecule, i.e., $X(4260)$ is more like a compact state. 

\subsection{$X_1 (2900)$}
In 2020, LHCb observed a spin-1 resonance around 2.9~GeV in the $D^- K^+$ invariant mass spectrum, labeled as $X_1 (2900)$~\cite{LHCb:2900}. This resonance is near the threshold of $\bar{D}_1 K$ and $\bar{D}^* K^*$, so it is more likely to be a hadronic molecule of these two hardron pairs. However, $X_1 (2900)$ has $J^P=1^-$ given by the experiment, and an $S$-wave $\bar{D}^* K^*$ molecule and $\bar{D}_1 K$ molecule has $J^P=1^+,~1^-$, respectively. So, if $X_1 (2900)$ is a hadronic molecule, it is more likely to be a molecule of $\bar{D}_1 K$.    

In Ref.~\cite{Chen:X2900}, two schemes of the production of $X_1 (2900)$ are taken into consideration. One is to produce $X_1 (2900)$ dynamically and another is to introduce $X_1 (2900)$ explicitly.  In these two schemes, channel $\bar{D}K$ and $\bar{D}_1 K$ are involved. The dynamically generated $X_1(2900)$ is achieved by the couple channel K-matrix formula,
\begin{equation}
    T=\mathcal{K}\cdot[1-i\rho(s)\mathcal{K}]^{-1},~\mathcal{K}=
\left(\begin{array}{cc}
\mathcal{T}_{\bar{D} K \to \bar{D} K} & \mathcal{T}_{ \bar{D}K\to \bar{D}_1 K}+\mathcal{P}_{12}\\
\mathcal{T}_{\bar{D}_1 K\to \bar{D}K}+\mathcal{P}_{12} & \mathcal{T}_{\bar{D}_1 K\to \bar{D}_1 K}
\end{array}\right)^{IJ}, 
\end{equation}
where $\rho=\mathrm{diag}\{\rho_{\bar{D}K} ,\rho_{\bar{D}_1 K}\}$, is the diagnoal matrix of phase space factors, $\mathcal{T}$s are the tree level diagrams with special isospin and angular momentum, $\mathcal{P}_{12}$ is a first order polynomial of c.m. energy to simulate the effects from higher order interactions such as vector meson exchanging diagrams in physical region. The explicitly introduced $X_1(2900)$ is achieved by the Flatt\'e-like parametrization.

The isospin symmetry is considered in the method of couple channel K-matrix, however, the values of parameter $\zeta_1$ which is involved in the tree-level amplitude $\mathcal{T}_{ \bar{D}K\to \bar{D}_1 K}$ are quite different in iso-singlet and iso-triplet solutions. This parameter comes from the interaction lagrangian~\cite{Ding:2008gr},
\begin{equation}
    \mathcal{L} = -i \zeta_{1}\left\langle\bar{H}_{a}^{(\bar{Q})}\left(\mathcal{V}_{\mu}-\rho_{\mu}\right)_{a b} T_{b}^{(\bar{Q}) \mu}\right\rangle+\mathrm{h.c.},
\end{equation}
where $\bar{H}^{(\bar{Q})}$ is the field operator with $J^P=0^-$ such as $\bar{D}$ and $T^{(\bar{Q})}$ is the field operator with $J^P = 1^+$ such as $\bar{D}_1$. $\rho_{\mu}$ is the vector meson octet. So, in principle, the parameter $\zeta_1$ should be estimated from the decay $D_1 \to D+\rho(\omega)$, but this process is forbidden kineically. In Ref.~\cite{Dong:Y(4260)}, it is estimated from the process $K_1\to K \rho$ and takes the value of about $\pm 0.16$~. Since the corresponding values of $\zeta_1$ given by the iso-triplet solution and iso-singlet solution are $5.85\pm 2.00$ and $-0.9\pm 0.48$, so it indicates that $X_1 (2900)$ is an iso-singlet.    

\begin{figure}[htpb]
    \centering
    \subfigure[]{
    \includegraphics[scale=0.2]{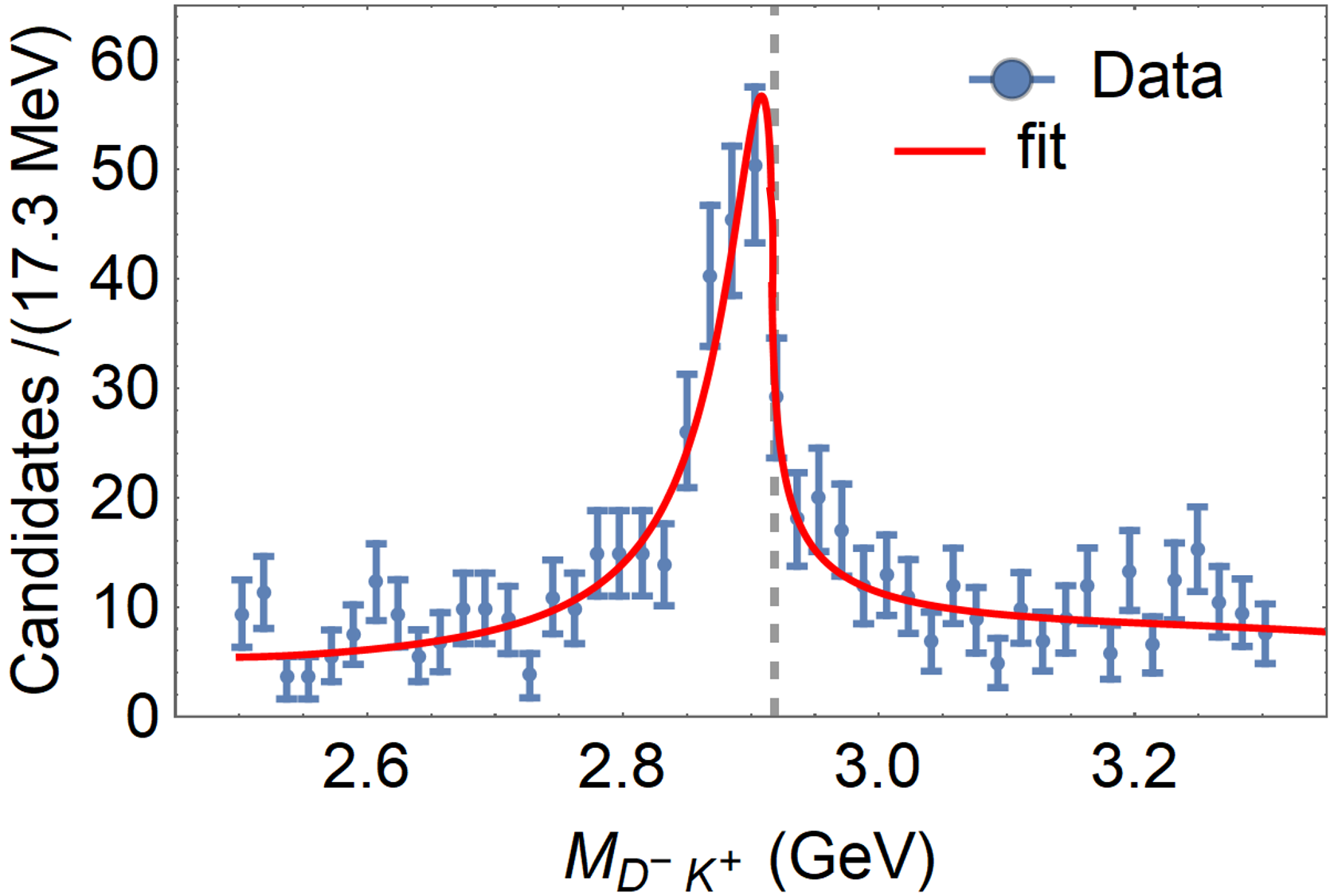}}
    \subfigure[]{
    \includegraphics[scale=0.55]{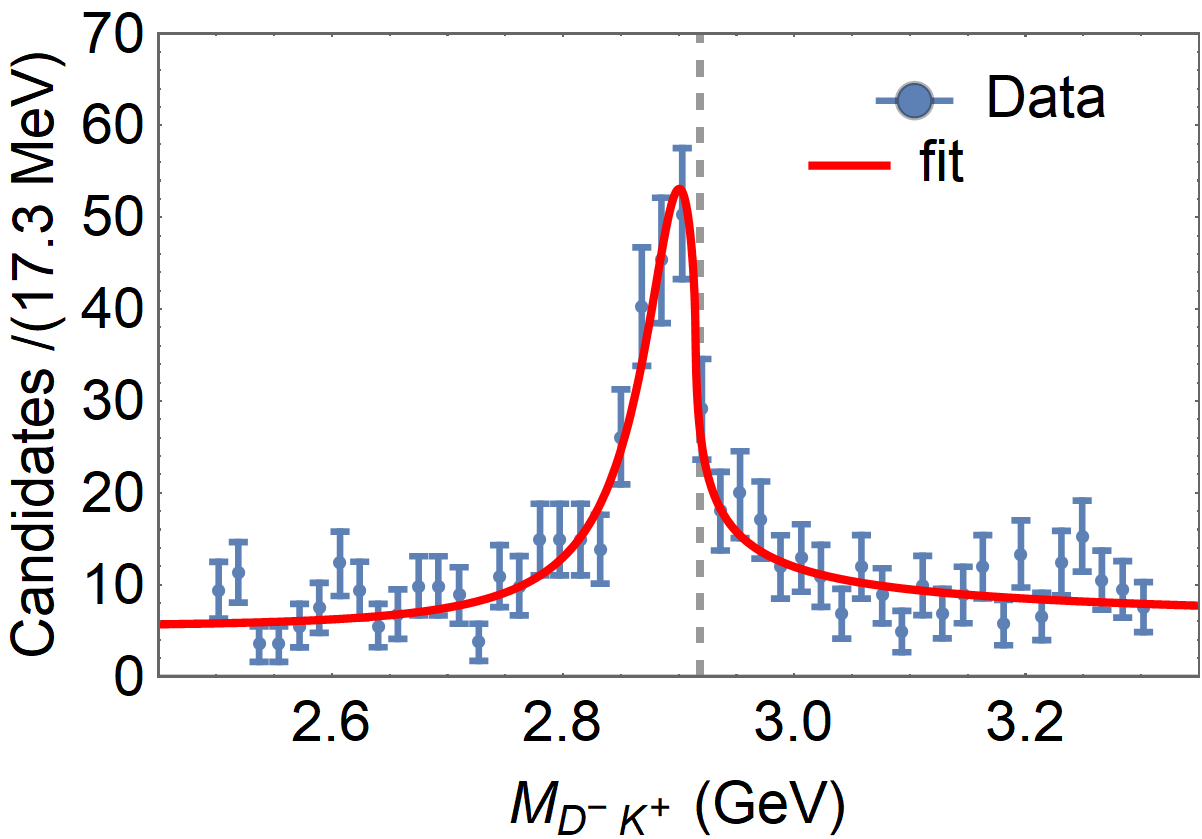}}
    \caption{Fit results in Ref.~\cite{Chen:X2900} for (a): dynamically generated $X_1(2900)$ with $IJ^P=01^-$, (b): explicitly introduced $X_1 (2900)$. These two fits have almost the same goodness and only one pole is found near the threshold of $\bar{D}_1 K$ in both cases.}
    \label{fig:X2900}
\end{figure}

In both two schemes, there is only one pole found near the threshold of $\bar{D}_1 K$. For dynamically generated $X_1(2900)$, the pole is located at $(2.928-0.034i)$~GeV and for explicitly introduced $X_1 (2900)$, the pole is located at $(2.910-0.039i)$~GeV. According to the PCR, this suggests that $X_1 (2900)$ is a hadronic molecule of $\bar{D}_1 K$ and at this time, two different schemes lead to the same conclusion, just like the study of $Z_c (3900)$ in Ref.~\cite{Gong:Zc3900_2}.

\section{Summary and conclusion}
 In this talk, it is showen that PCR provides a model-independent method to distinguish hadronic molecules from compact states. But it still has a disadvantage that only $S$-wave amplitudes can be used as inputs. Nevertheless, almost all the known molecular states are $S$-wave couplings, so this method is well practical. 

In this talk, the applications of PCR in studying some XYZ states such as $X(3872),~Z_c(3900),$ $X(4660),~X(6900),~X_1(2900)$ are reviewed. With the help of PCR, $X(3872)$ is regarded as charmoniums, i.e., compact states and $Z_c (3900)$, $X_1 (2900)$ are suggested to be hadronic molecules of $\bar{D}^* D$ and $\bar{D}_1 K$, respectively. For $X(6900)$, its nature cannot be well determined using current data, so more experimental information are required. Furthermore, it is mentionable that in some works, molecule solutions for $X(3872)$ can be found, e.g., in Ref.~\cite{Guo:molecule} where it is argued that $\bar{D}^* D$ molecule. However, the analysis for the decay width to lower channels $\Gamma_c$ and the $B$ decay branch ratio $\mathcal{B}(B \to X(3872)K)$ as shown in Refs.~\cite{Meng:X3872,Zhang:X3872} suggest that $X(3872)$ is more like a charmonium despite the strong coupling to $\bar{D}^* D$.

From the applications listed in this paper, it is suggested that all the XYZ states which have exotic quantum number are more likely to be hadronic molecules and the state like $X(3872)$, is compact states since they have $\bar{c}c$ quantum numbers. For more discussion about methods in studying exotic hadron states, one is referred to the review~\cite{Yao:2020bxx}. 

\section*{Acknowledgements}
The authors would like to thank Han-Qing Zheng and Qin-Fang Cao for helpful discussions. This work is supported
in part by National Nature Science Foundations of China (NSFC) under contract numbers 11975028 and 10925522.

\end{document}